\documentclass[letterpaper]{article} 
\usepackage{aaai2026}  
\usepackage{times}  
\usepackage{helvet}  
\usepackage{courier}  
\usepackage[hyphens]{url}  
\usepackage{graphicx} 
\usepackage{natbib}  
\usepackage{caption} 
\usepackage{algorithm}
\usepackage{algorithmic}
\usepackage{multirow}
\usepackage{subfig}
\usepackage{amsmath} 

\usepackage{amssymb}
\usepackage{graphicx}
\usepackage{subcaption}
\usepackage{multirow}
\usepackage[table,xcdraw]{xcolor}

\usepackage{newfloat}
\usepackage{listings}
\DeclareCaptionStyle{ruled}{labelfont=normalfont,labelsep=colon,strut=off} 
\floatstyle{ruled}
\newfloat{listing}{tb}{lst}{}
\floatname{listing}{Listing}
\title{AAAI Press Formatting Instructions \\for Authors Using \LaTeX{} --- A Guide}

\title{Bid Farewell to Seesaw: Towards Accurate Long-tail Session-based Recommendation via Dual Constraints of Hybrid Intents}
\author {
    Xiao Wang,
    Ke Qin,
    Dongyang Zhang,
    Xiurui Xie,
    Shuang Liang\footnote{Corresponding author.}
}
\affiliations {
    University of Electronic Science and Technology of China\\
    wangxiao16@std.uestc.edu.cn, qinke@uestc.edu.cn,
    dyzhang@uestc.edu.cn,
    xiexiurui@uestc.edu.cn,
    shuangliang@uestc.edu.cn
}

\usepackage{bibentry}

\begin{document}

\maketitle

\begin{abstract}
Session-based recommendation (SBR) aims to predict anonymous users' next interaction based on their interaction sessions. In the practical recommendation scenario, low-exposure items constitute the majority of interactions, creating a long-tail distribution that severely compromises recommendation diversity. Existing approaches attempt to address this issue by promoting tail items but incur accuracy degradation, exhibiting a "see-saw" effect between long-tail and accuracy performance. We attribute such conflict to session-irrelevant noise within the tail items, which existing long-tail approaches fail to identify and constrain effectively. To resolve this fundamental conflict, we propose \textbf{HID} (\textbf{H}ybrid \textbf{I}ntent-based \textbf{D}ual Constraint Framework), a plug-and-play framework that transforms the conventional "see-saw" into "win-win" through introducing the hybrid intent-based dual constraints for both long-tail and accuracy. Two key innovations are incorporated in this framework: (i) \textit{Hybrid Intent Learning}, where we reformulate the intent extraction strategies by employing attribute-aware spectral clustering to reconstruct the item-to-intent mapping. Furthermore, discrimination of session-irrelevant noise is achieved through the assignment of the target and noise intents to each session. (ii) \textit{Intent Constraint Loss},  which incorporates two novel constraint paradigms regarding the \textit{diversity} and \textit{accuracy} to regulate the representation learning process of both items and sessions. These two objectives are unified into a single training loss through rigorous theoretical derivation. Extensive experiments across multiple SBR models and datasets demonstrate that HID can enhance both long-tail performance and recommendation accuracy, establishing new state-of-the-art performance in long-tail recommender systems. The source code is available at "https://github.com/jarviswww/Code4HID".
\end{abstract}


\section{Introduction}

\begin{figure}[t]
\centering
\includegraphics[width=3.2in]{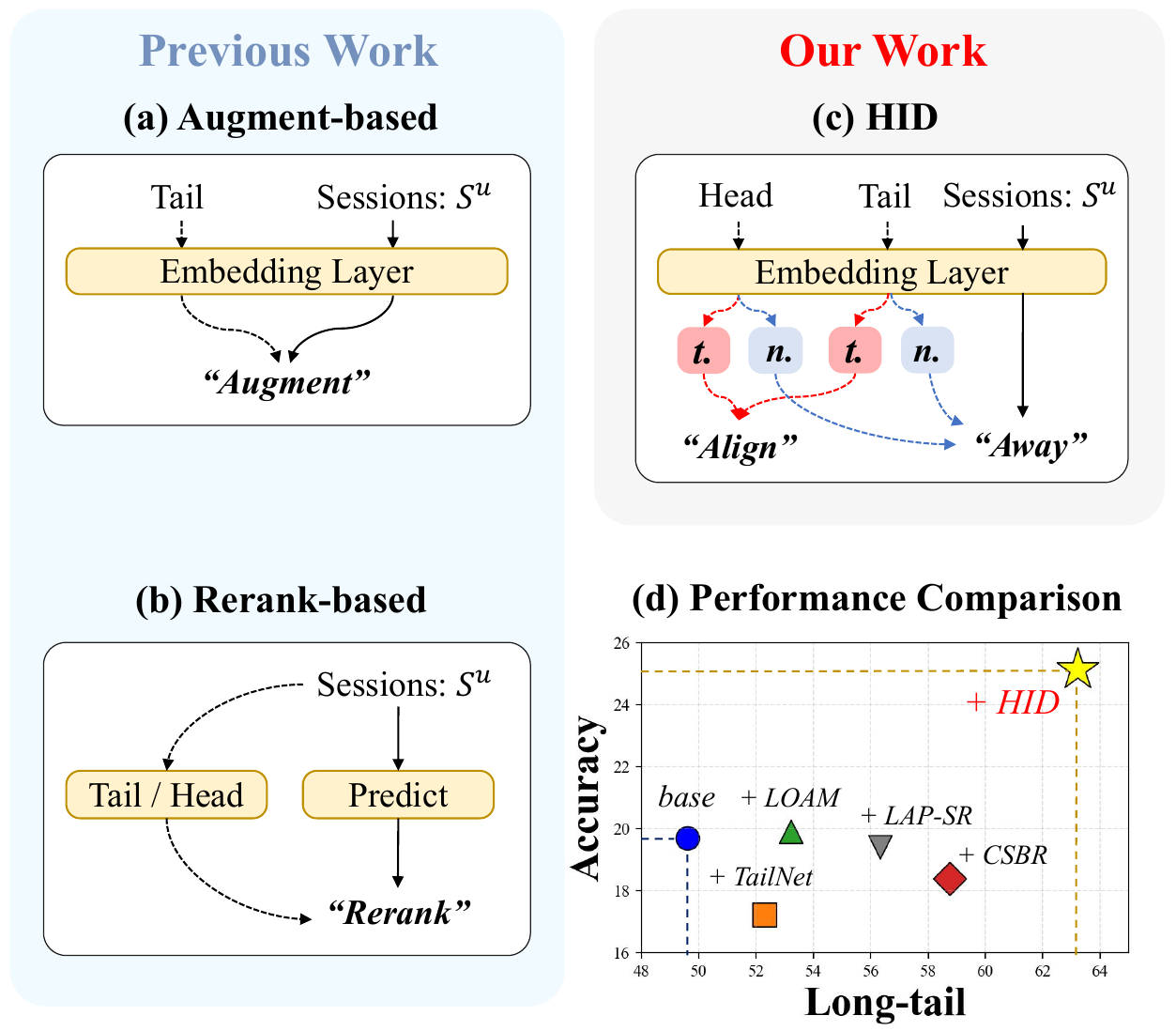}
\caption{Comparison between the our proposed HID and previous work.  (a) illustrates the design of HID, where \textit{\textbf{t.}} and \textit{\textbf{n.}} denotes target and noise items for session $S^u$, respectively; (c) and (d) demonstrate the frameworks of previous long-tail approaches. (b) evaluates the accuracy (i.e., HR@20) and long-tail performance (i.e., tCov@20) of the base SBR model GRU4Rec~\cite{Hidasi16gru4rec} and GRU4Rec + long-tail approaches on Tmall dataset.}    
\vspace{-0.4cm}
\label{Fig: Pre_Our}
\end{figure}

\begin{figure}[t]
\centering
\includegraphics[width=3.2in]{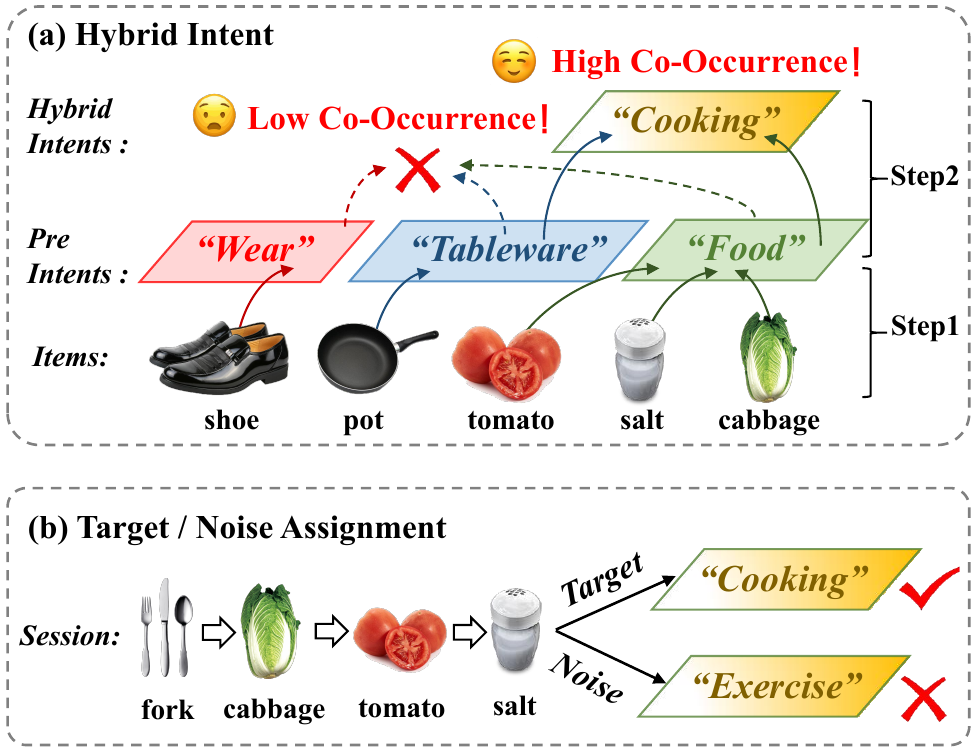}
\caption{The demonstration of: (a) Hybrid Intent: Step 1 groups items by shared attributes (e.g., food) as the \textit{preliminary intents}; Step 2 combines attributes with high co-occurrence (e.g., food + pot) to form the \textit{hybrid intents} (e.g., cooking). (b) 
        Intent Assignment: Assigns target (relevant) and noise (irrelevant) hybrid intents to anonymous sessions.}    
\vspace{-0.4cm}
\label{Fig: HybridIntent}
\end{figure}

Session-based recommendation (SBR) addresses information overload by predicting the next item from short-term interactions, particularly in privacy-sensitive scenarios lacking long-term user profiles~\cite{Li2025Survey, Latifi21Session}. While deep learning methods in SBR (e.g., deep sequential models~\cite{Hidasi16gru4rec, Li17Narm, liu18Stamp, yuan21Dsan, Hou22Core} and deep graphic models~\cite{Wu19Srgnn, Qiu19Rethinking, Wang20Gcegnn, Xia21Dhcn, Xia21Cotrec, Pan20Star}) can effectively model item correlations, their model-centric focus overlooks inherent data biases. A key challenge is the long-tail distribution in recommendation data~\cite{Sundaresan2011Recommender, Yang2023LOAM, Liu2020TailNet}, where a small number of high-exposure items (i.e., head items) dominate the model's attention, while a significantly larger number of low-exposure items (i.e., tail items) are often disregarded. This unfair phenomenon leads to the overlooking of potentially essential but low-exposure tail items, limiting the diversity of recommendations~\cite{Turgut2023Diversity, Yin2024Diversity, Lee2024PostTraining}. Besides, the long-tail distribution causes the model to be more inclined to recommend head items, resulting in a vicious cycle. 

Previous advancements in long-tail SBR focus on developing plugins that seamlessly integrate with existing SBR models, emphasizing the significance of tail items to mitigate the long-tail issue~\cite{Liu2020TailNet, Chen2023CSBR, Yang2023LOAM, Peng2024LAPSR}. Broadly, long-tail SBR approaches fall into two categories: (i) Augment-based approaches, which employ augmentation strategies to refine the tail item embeddings or session embeddings~\cite{Yang2023LOAM, Kim2023MELT, Huang2024CSLP, Liu24LLMESR}, and (ii) Rerank-based approaches, which predict head/tail item distributions based on interaction sessions and directly modify the final ranking results~\cite{Liu2020TailNet, Chen2023CSBR, Peng2024LAPSR}. Both approaches consistently emphasize the significance of tail items. The brief demonstrations of their frameworks are given in (a) and (b) of Figure~\ref{Fig: Pre_Our}. Despite their success, two critical limitations remain unresolved: (i) their undifferentiated emphasis on tail items introduces session-irrelevant noise (e.g., "clothing" for a session consists of books.), as \textit{\textbf{not all tail items align with session-specific user requirements}}, resulting in the degradation of recommendation accuracy, and (ii) they lack explicit supervisory signals for long-tail objectives, thus relying on indirect optimization via cross-entropy loss. Crucially, such augmentation and re-ranking strategies often conflict with the cross-entropy optimization objective due to the inclusion of potential session-irrelevant items, resulting in a “see-saw” effect~\cite{Wang2021Denoising, Wei2024LLMRec}. To address these flaws, our work revolves around two key innovations: (i) the effective discrimination of noise, restricting the consideration of long-tail issues to session-relevant tail items, and (ii) the introduction of explicit long-tail supervisory signals to concurrently improve the long-tail and accuracy performance. 

For the noise discrimination, given that interaction sessions are driven by user intent~\cite{Li23Intention-aware, Wang2024HearInt}, we employ intent modeling to capture the overarching preference of the anonymous user. Previous work primarily derives user intents from restricted sequential segments (e.g., sliding windows) or semantically clustered items within individual sessions~\cite{Wang19Modeling, Zhang23Efficiently,Choi2024MiaSRec, Wang2024HearInt}, but suffer from unreliable intent extraction due to noise interference and neglect cross-session intent consistency~\cite{Choi2024Intent,Wang2024HearInt}. Therefore, we
propose the \textit{hybrid intent}, which captures the user preference through attribute consistency (e.g., commodity categories, music genres) and item co-occurrence patterns, as shown in (a) of Figure~\ref{Fig: HybridIntent}. Following this, we assign target and noise intents to each session to enable the discrimination of session-irrelevant items, as shown in (b) of Figure~\ref{Fig: HybridIntent}.

For the long-tail supervisory signals, since the long-tail issue stems from the disparity in embedding distributions between head and tail items, resulting in discrepancies in their similarity to user embeddings~\cite{Yin2012LongTail, Gupta19NISER}, we propose explicit constrains on these similarities during the training process to provide direct supervised signal. Specifically, to address the distribution inconsistency between head and tail items, we align their similarity scores to each session through a novel constraint objective. This constraint is termed the \textit{Constraint for Long-tail}, which operates exclusively on session-relevant items (i.e., items belong to the target intents). Furthermore, to ensure the recommendation accuracy, we introduce an additional \textit{Constraint for Accuracy} that explicitly enlarges the similarity discrepancy between sessions and session-irrelevant items (i.e., items belong to noise intents) during the training process. The mutual independence of target and noise intents ensures that the two constraints are not conflicting. The brief framework of constraints is given in Figure~\ref{Fig: Pre_Our} (c).

Incorporating the above innovations, we name this novel approach as the \textbf{H}ybrid \textbf{I}ntent-based \textbf{D}ual Constraint Framework (\textbf{HID}). This model-agnostic and plug-and-play framework can be easily integrated into existing SBR models. Specifically, HID consists of the \textit{hybird intent learning} module and the \textit{intent constraint loss} (ICLoss). The \textit{hybird intent learning} module first aggregates items that share the same attribute to form preliminary intent units. Subsequently, based on the attribute co-occurrence relations from all interaction sessions, a \textit{preliminary intent graph} is constructed whose nodes are the preliminary intents and edge weights represent their co-occurrence frequency. After that, we employ spectral clustering, grouping the preliminary intents into hybrid intents. Furthermore, we derive the theoretical formulations of the \textit{Constraint for Long-tail} and the \textit{Constraint for Accuracy}, and combine them to acquire the \textit{intent constraint loss}, which aligns embeddings of head and tail items within the target intent while repelling noise intents from the current session in the feature space. As shown in (d) of Figure~\ref{Fig: Pre_Our}, HID achieves significant improvements in both accuracy and diversity over previous long-tail competitors, due to its session-irrelevant noise discrimination capability and dual constraints of long-tail and accuracy.

To sum up, we conclude the main contributions of this work as follows:

\begin{itemize}
\item We propose a novel framework named HID aimed at achieving accurate long-tail SBR. Its brevity ensures easy reproduction and integration with existing SBR models.
\item We innovatively propose a novel concept of the hybrid intent, which advances session-based recommendation by jointly modeling attribute-level correlations and attribute co-occurrence patterns to redefine the item-intent mapping.

\item We explicitly model the learning objective of accurate long-tail SBR through two novel constraint paradigms for both the long-tail and accuracy, and integrate them into a unified, theoretically-grounded intent constraint loss that optimizes both objectives.
\item Extensive experiments conducted on various SBR models and long-tail competitors demonstrate the effectiveness of HID in addressing the long-tail issue and improving recommendation accuracy.
\end{itemize}

\section{Related Works}

\textbf{Augment-based Approaches}. This technical route primarily focuses on enhancing the embeddings of tail items or emphsizing the significance of tail items when generating the session embeddings. 
LOAM~\cite{Yang2023LOAM} enhances tail items and sessions through the Niche-Walk Augmentation and Tail Session Mixup. GALORE~\cite{Luo2023Improving} introduces a graph augmentation approach to enhance the edge of tail items in the interaction graph. GUME~\cite{Lin2024GUME} employs the graphs and user modalities enhancement. MelT~\cite{Kim2023MELT} employs mutual enhancement of tail users and items, which jointly mitigates the long-tail issue. Additionally, some approaches have explored the useage of large language models (LLMs). LLM-ESR~\cite{Liu24LLMESR} utilizes the semantic embeddings derived from LLMs to enhance the tail items.

\textbf{Rerank-based Approaches}. This technical route aims to infer the distribution of tail and head items from interaction sessions, thereby enabling direct adjustment of recommendation results. TailNet~\cite{Liu2020TailNet} introduces a preference mechanism to predict the adjustment index of head and tail items. CSBR~\cite{Chen2023CSBR} proposes two additional training objectives: distribution prediction and distribution alignment to calibrate the recommendation results. LAP-SR~\cite{Peng2024LAPSR} 
adjusts the weight scores of recommended items based on the long-tail items and the intra-session similarity.

Although the above methods have made contributions to addressing the long-tail problem, they all neglect the consideration of noise in tail items and lack explicit modeling of the long-tail objective.

\section{Preliminaries}
\subsection{Problem Definition}
\label{Pre: Problem Definition}

Let $V = \{v_1, v_2, \dots, v_m\}$ represent the set of all unique items, where $m$ is their total counts. An anonymous session is represented as $S^u = \{v^u_1, v^u_2, \dots, v^u_l\}$, where $u$ is the session ID, $l$ is the length of the interaction session, and $v^u_t \in V$ (0 < t < $l$) is the item ID which is interacted at timestep $t$. \textit{In this paper, all symbols in bold represent the vector embeddings.} For example, in $S^u = \{\textbf{v}^u_1, \textbf{v}^u_2, \dots, \textbf{v}^u_l\}$, $\textbf{v}^u_t \in \mathbb{R}^d$ represents the vector embedding of item $v^u_t$. Given a session $S^u$, the task in session-based recommendation is to predict the next-interacted item $v^u_{l+1}$ (i.e., the ground truth item). According to the Pareto principle~\cite{Bos1986Analysis}, the top 20\% of items with the highest frequency of occurrence are considered to be head items, while the remainings are tail items.

\subsection{Session-based Recommendation Models}
\label{Sect: Preliminaries-Session-based Recommendation Models}
Session-based recommendation (SBR) models follow a two-stage paradigm: a SBR encoder to transform the inputs into session embeddings, and a prediction layer to generate the recommendations. The basic structure of the SBR model is demonstrated in the blue components of Figure~\ref{Fig: model}.

Given a session $S^u = \{\textbf{v}^u_1, \textbf{v}^u_2, \dots, \textbf{v}^u_l\}$, whose vector embeddings are initialized using the Gaussian distribution, SBR models typically propagate it into a SBR encoder, which is denoted as $F(x)$ in Figure~\ref{Fig: model}, to generate the session embedding: $\textbf{S}^u =  F(S^u)$,
where $S^u \in \mathbb{R}^{l\times d}$, $\textbf{S}^u \in \mathbb{R}^{d}$. 

After acquiring session embedding $\textbf{S}^u$, SBR models multiply it with the candidate item embeddings and apply a softmax to calculate the probabilities of each item being the next-interacted one: $y'_i = \text{softmax}({\textbf{S}^u}^T {\textbf{v}_i})$, where $\textbf{v}_i$ is the embedding of item $v_i \in V$. Then, the next-item prediction task is adopted as the learning objective, where the cross-entropy loss is usually leveraged as the objective function: $\mathcal{L}_p = - \sum^{m}_{i=1} y_i \text{log}(y'_i)$. where $y_i$ is the one-hot encoding vector of the ground truth.

\section{Proposed Method}
\label{Sect: Proposed Method}

\subsection{Hybrid Intent Learning}
\label{Sect: Proposed Method-Hierarchical Clustering}
Existing intent mining approaches exhibit two weaknesses: (i) only temporal relations among items are considered, which is not always reliable due to the interaction noise, and (ii) only a single session is considered, neglecting that items from different sessions can reflect the same intent. Therefore, we propose attribute-aware spectral clustering, giving the brief demonstration in right part of Figure~\ref{Fig: model}. 

Note that the whole process of acquiring hybrid intents can be pre-computed and stored locally. Therefore, during training or serving, only providing the item for retrieval enables the acquisition of hybrid intents.

\subsubsection{Preliminary Intent.}
Since items sharing the same attribute can typically reflect similar user preferences (e.g., electronic products or books), we consider the item attribute as the preliminary intent unit. Given item attribute set $C' = \{c'_1, c'_2,...,c'_{k}\}$, where $c'_i$ (1<i<$k$) is the $i\text{-}th$ attributes that represents a specific preliminary intent, and $k$ is their total counts. For each $c'_i$, we denote it as a set of items $c'_i = \{v_{c_i,1},v_{c_i,2},..., v_{c_i,|c'_i|}\}$.


\subsubsection{Preliminary Intent Graph.}
To explore the attributes relations within all sessions, we first replace the item IDs within each session with their corresponding attribute IDs. After that, we iterate over all attributes within each session and count the 1-hop neighbors of each attribute, along with the frequency of their occurrences to form the preliminary intent graph. This intent graph is denoted as $\mathcal{G}=(\mathcal{P}, \mathcal{E}, \mathcal{W})$, where $\mathcal{P}$ is the set of attribute IDs, $\mathcal{E} = \{(c'_i,c'_j)~|~c'_i \in C', c'_j \in \mathcal{N}_{c'_i}\}$ is the edge between attribute $c'_i$ and $c'_j$, where $\mathcal{N}_{c'_i}$ is the neighbor set of attribute $c'_i$, and $\mathcal{W}$ is the set of weights, where $w_{ij} \in \mathcal{W}$ of the edge $(c'_i, c'_j)$ is the co-occurrence frequency of attribute $c'_i$ and $c'_j$.

\begin{figure}[t]
\includegraphics[width=3.2in]{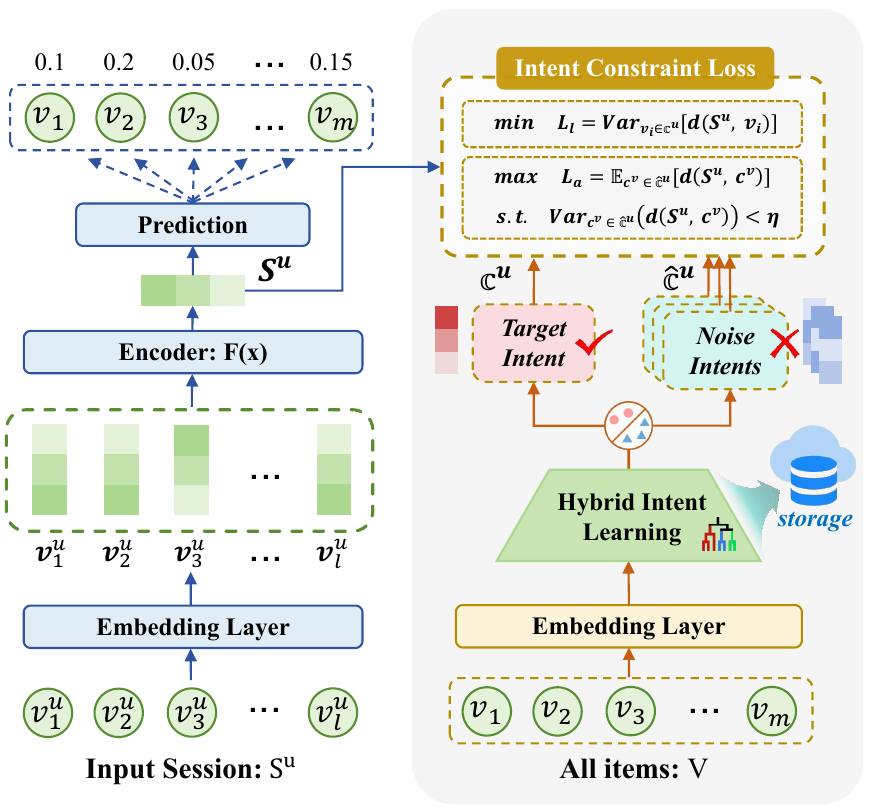}
\caption{The overall architecture of $SBR~model$ (left) + $HID$ (right). The Hybrid Intent Learning module first assigns items to $k$ preliminary intents, and then further divides them into $n$ hybrid intents $C$ based on the topological relationships in the preliminary intent graph. After refining the hybrid intents, the intent constraint loss is introduced to regulate the learning process of session embedding $\textbf{S}^u$.}    
\vspace{-0.4cm}
\label{Fig: model}
\end{figure}

\subsubsection{Hybrid Intent.}
After acquiring the preliminary intent graph $\mathcal{G}$, with the aim of mining the global co-occurance patterns of attributes, the spectral clustering is employed to learn the topological relations among attributes. Given the graph $\mathcal{G}=(\mathcal{P}, \mathcal{E}, \mathcal{W})$, we first calculate its Laplace matrix:
\begin{align}
    L = I - D^{-\frac{1}{2}}WD^{-\frac{1}{2}},
\label{Equ: Laplace Matrix}
\end{align}
where $D_{ii} = \sum_j w_{ij}$. Then, we compute the eigenvalues and eigenvectors of the normalized Laplacian matrix $L$. Let $\lambda_1 \leq \lambda_2\leq ... \leq \lambda_q$ be the smallest $q$ eigenvalues and their corresponding eigenvectors $\hat{\lambda}_1, \hat{\lambda}_2,..., \hat{\lambda}_q$ form the eigenvector matrix. Each row of eigenvector matrix represents the embedding of a node in the reduced q-dimensional space. 

After that, we apply the k-means algorithm on the rows of the eigenvector matrix. The $i\text{-th}$ row of the eigenvector matrix corresponds to the $i\text{-th}$ attribute of $C'$, which also corresponds to a node in the preliminary intent graph. Therefore, the attributes are reclassified into $n$ clusters. Since the attributes reprsent the preliminary intents, we combine attributes belonging to the same cluster to form the hybrid intent. The hybrid intent set is defined as $C = \{c_1,c_2,...,c_n\}$. The embedding of the hybrid intent is derived from the item embeddings associated with the attributes it contains. To reduce time complexity, we concatenate the items within attributes and then apply average pooling to obtain the hybrid intent embedding, which can be formulated as follows:
\begin{align}
    \textbf{c}_i = \frac{1}{|c_i|}\underset{v_{j} \in c_i}{\sum} \textbf{v}_{j}.
\label{Equ: Hybrid Intent Embedding}
\end{align}
where $ c_i = \{v_{c_i,1},v_{c_i,2},...,v_{c_i,|c_i|}\}$, $v_{c_i,1}$ to $v_{c_i,|c_i|}$ are the items from the attributes that form the hybrid intents $c_i$, and $1<i<n$.

\subsubsection{Target and Noise Intents.}
After acquiring the set of hybrid intents, for each batch of sessions $\mathcal{B} = \{S^1, S^2,..., S^b\}$, we define the \textit{target intent} and \textit{noise intents} for session $S^u = \{v^u_1, v^u_2,..., v^u_l\}$ where $1<u<b$ as follows:

\textbf{Definition 1} (Target Intent). \textit{For session $S^u$, the hybrid intents that contain its next-item $v^u_{l+1}$ are considered as its \textit{target intent} set $\mathcal{C}^{u}$.}
\begin{align}
    \mathcal{C}^{u} = \{c_i~|~v^u_{l+1} \in c_i, c_i \in C\}
\end{align}
\textbf{Definition 2} (Noise Intent). \textit{For session $S^u$, given the minibatch $\mathcal{B}$, target intents of other sessions $S^{v} \in \mathcal{B} \setminus S^{u}$ that are not within $\mathcal{C}^{u}$ are considered as its noise intent set $\mathcal{\hat{C}}^{u}$.}
\begin{align}
    \mathcal{\hat{C}}^{u} =\{c_i~|~v^{v}_{l+1} \in c_i, S^{v} \in \mathcal{B} \setminus S^{u}, c_i \in C\setminus \mathcal{C}^{u}\}
\end{align}
Both the target and noise intents are only leveraged in the intent constraint loss as supervisory signals during the training stage, so there is no risk of data leakage.

\subsection{Dual Constraints for Long-tail and Accuracy}
\label{Sect: Proposed Method-Intent Constraint Loss}

Following the extraction of hybrid intent embeddings $\textbf{C} = \{\textbf{c}_1,\textbf{c}_2,..., \textbf{c}_n\}$,  the subsequent objective involves imposing constraints on the learning process of session embeddings. Given the session embedding $\textbf{S}^u$ learned by traditional SBR models such as STAMP~\cite{liu18Stamp} or SRGNN~\cite{Wu19Srgnn}, our next aim is to construct supervisory signals regarding the long-tail performance and recommendation accuracy. The demonstration of these supervisory signals are given in Figure~\ref{Fig: model}.

The sequential nonlinear transformations in $F(x)$ introduces potential misalignment between the scale of hybrid intent embeddings and derived session embeddings. To mitigate this discrepancy and enforce commensurable embedding spaces, we employ $L_2$-norm to project both embedding sets onto a unit hypersphere, thereby establishing a unified metric space for subsequent operations: 
\begin{align}
    \textbf{c}_i = \frac{\textbf{c}_i}{||\textbf{c}_i||_2}, \
    \textbf{S}^u = \frac{\textbf{S}^u}{||\textbf{S}^u||_2}.
\label{Equ: L2 Norm}
\end{align}
Subsequently, we delineate the implementation details of the \textit{Constraint for Long-tail} and the \textit{Constraint for Accuracy}, introducing their formulations and roles in the optimization framework.

\subsubsection{Constraint for Long-tail} 
\label{Sect: Proposed Method-Intent Constraint Loss-Constraint for Long-tail}
The long-tail problem emerges due to the pronounced disparity in session-item similarity between tail and head items, as documented in previous research~\cite{Yin2012LongTail, Liu20Longtail}. Based on this observation, we propose a novel constraint: minimizing the variance of similarity scores between sessions and items belonging to the target intent. This constraint can reduce the divergence in similarity distributions between session-to-head and session-to-tail, thereby promoting more balanced recommendation performance. Formally, the constraint is defined as follows:

\textbf{Definition 3} (Constraint for Long-tail). \textit{Given the session embedding $\textbf{S}^u$, the variance of its Euclidean distances to the embeddings of all items belonging to its \textit{target intent} should be minimized, which can be formulated as:}
\begin{align}
    \min~\mathcal{L}_l = \text{Var}_{v_i \in \mathcal{C}^u}\left[d(\textbf{S}^u, \textbf{v}_i)\right].
\label{Equ: CoL1}
\end{align}
where Var is the variance calculation, and $d(x,y)$ measures the Euclidean distance between x and y. The time complexity of the above operation is $O(Nd)$, where N is the number of items belonging to the target intent $\mathcal{C}^u$, and d is the embedding dimension. Since HID is a model-agnostic plugin, the complexity is a key concern. Therefore, we further propose an approximate formulation of Equation~(\ref{Equ: CoL1}) with lower complexity. As shown in the following theorem:





\textbf{Theorem 1} (Optimizing Equivalence). \textit{The Equation~(\ref{Equ: CoL1}) with time complexity of $O(Nd)$ can be approximated to an equation with time complexity of $O(d)$ during the optimization process:}
\begin{align}
    &\min~\mathcal{L}_l =\text{Var}_{v_i \in \mathcal{C}^u}[d(\textbf{S}^u, \textbf{v}_i)]~~\sim ~~\min~d(\textbf{S}^u, \textbf{c}^u).
\label{Equ: CoL2}
\end{align}
where $\textbf{c}^u$ is the embedding of the target intent. The detailed proof of Theorem 1 is provided in \textbf{Appendix~A}. Given the Theorem 1, the optimization process that maximizes the similarity between the session embedding $\textbf{S}^u$ and the target intent embedding $\textbf{c}^u$ is mathematically equivalent to solving Equation~(\ref{Equ: CoL1}). This concise constraint provides an efficient mechanism for enhancing tail item coverage within the target intent space while excluding noise intents.

\subsubsection{Constraint for Accuracy} 
\label{Sect: Proposed Method-Intent Constraint Loss-Constraint for Accuracy}
To further mitigate session-irrelevant recommendations, it is crucial to proactively limit the presence of noise items in the recommendations. Therefore, we propose minimizing the mean of similarity scores between sessions and noise intents. Besides, to prevent extreme cases, the variance of similarity scores should also be constrained. By regulating both the mean and variance, we ensure that the noise intent distribution remains distant from the specific sessions. This constraint can be formulated as:

\textbf{Definition 4} (Constraint for Accuracy). \textit{Given the session representation $\textbf{S}^u$, the mean and vairance of its Euclidean distances to the representations of \textit{noise intents} within the same batch should be maximized and restricted, respectively, which can be formulated as:}
\begin{equation}
    \begin{split}
    \max~~ &\mathcal{L}_{a} = \mathbb{E}_{c^{v} \in \mathcal{\hat{C}}^u}d(\textbf{S}^u, \textbf{c}^{v}) \propto \sum_{c^{v} \in \mathcal{\hat{C}}^u}d(\textbf{S}^u, \textbf{c}^{v}),~~\\  
    \text{s.t.}~~&\text{Var}_{c^{v} \in \mathcal{\hat{C}}^u}\left(d(\textbf{S}^u, \textbf{c}^{v})\right) < \eta,
    \end{split}
\label{Equ: CoA2}
\end{equation}
where $\eta$ is the threshold of variance.

\subsubsection{Intent Constraint Loss} 
Optimizing these two constraints independently presents certain challenges. Therefore, we combine the objectives of Equation~(\ref{Equ: CoL2}) and Equation~(\ref{Equ: CoA2}), unifying them into a single loss function:
\begin{equation} 
    \begin{split}
    \min~~&\mathcal{L}_c = \sum_{S^u \in \mathcal{B}}\text{log}\frac{\text{exp}(d(\textbf{S}^u, \textbf{c}^u))}{\text{exp}(d(\textbf{S}^u, \textbf{c}^u))+\underset{c^{v} \in \mathcal{\hat{C}}^u}{\sum}\text{exp}(d(\textbf{S}^u, \textbf{c}^v))}, ~~\\
    \text{s.t.}~~&\text{Var}_{c^{v} \in \mathcal{\hat{C}}^u}(d(\textbf{S}^u, \textbf{c}^{v})) < \eta,
    \end{split}
\label{Equ: ICLoss-1}
\end{equation}
where exp(x) is leveraged to amplify the difference between the target and noise intents, $\text{exp}(d(\textbf{S}^u, \textbf{c}^u))$ is incorporated into the denominator to stabilize the loss range and mitigate the impact of the number of negative samples on the loss scale. To further minimize the effect of the noise intents, we give another theorem:

\textbf{Theorem 2} (Triplet Loss Approximation). \textit{The optimization of the objective function in Equation~(\ref{Equ: ICLoss-1}) is approximately proportional to optimize a (N-1)-triplet loss with a fixed margin of 2}:
\begin{align}
    \begin{split}
    &\mathcal{L}_c \propto \sum_{S^u \in \mathcal{B}} \sum_{c^{v} \in\mathcal{\hat{C}}^u}\left(\|\textbf{S}^u-\textbf{c}^u\|^{2}-\|\textbf{S}^u-\textbf{c}^v\|^{2}+\textbf{2}\right).
    \end{split}
\label{Equ: (N-1)TupletLoss)}
\end{align}
The proof of Theorem 2 is given in \textbf{Appendix~B}. The constant term '2' is the fixed margin that decides the distinction of $d(\textbf{S}^u, \textbf{c}^u)$ and $d(\textbf{S}^u, \textbf{c}^v)$. However, this fixed margin is inadequate for distinguishing the target and noise intents, especially in scenarios with high variability in intent distributions or in the presence of ambiguous intents. Therefore, we introduce a flexible coefficient to replace the original constant, enabling flexible margin adjustment based on the recommendation scenario: 
\begin{align}
    &\min~~\mathcal{L}_c = \\ &\sum_{S^u \in \mathcal{B}}\text{log}\frac{\text{exp}(d(\textbf{S}^u, \textbf{c}^u)/\sigma )}{\text{exp}(d(\textbf{S}^u, \textbf{c}^u)/\sigma )+\underset{c^{v} \in\mathcal{\hat{C}}^u}{\sum}\text{exp}(d(\textbf{S}^u, \textbf{c}^v)/\sigma )},\\~~\text{s.t.}~~&\text{Var}_{c^{v} \in\mathcal{\hat{C}}^u}(d(\textbf{S}^u, \textbf{c}^{v})) < \eta,
\label{Equ: ICLoss-2}
\end{align}
where $\sigma$ is the flexible coefficient. To directly apply the gradient descent for updates and avoid the complexity of constraint optimization, we reformulate the hard variance constraint $\text{Var}_{c^{v} \in \mathcal{\hat{C}}^u}(d(\textbf{S}^u, \textbf{c}^{v})) < \eta$ as a penalty term $p^u$:
\begin{align}
    p^u=\max (0, \underset{c^{v} \in \mathcal{\hat{C}}^u}{\text{Var}}(d(\textbf{S}^u, \textbf{c}^{v}))-\eta).
\label{Equ: Penalty}
\end{align}
In addition, previous research has found that cosine similarity can achieve better alignment and uniformity of embeddings~\cite{Wang2020Understanding}. Therefore, we adopt cosine similarity instead of Euclidean distance.
The final training objective of the intent constraint loss (ICLoss) is formulated as:
\begin{equation}
    \begin{split}
    \min~~\mathcal{L}_c = -\sum_{S^u \in \mathcal{B}}\text{log}\frac{\mathbf{X}}{(1+\lambda p^u)(\mathbf{X}+\mathbf{Y})},
    \end{split}
\label{Equ: ICLoss-3}
\end{equation}
where $\lambda$ is the hyper-parameter that controls penalty, $\mathbf{X}$ is $\text{exp}(\text{cos}(\textbf{S}^u, \textbf{c}^u)/\sigma)$, $\mathbf{Y}$ is $\underset{c^{v} \in  \mathcal{\hat{C}}^u}{\sum}\text{exp}(\text{cos}(\textbf{S}^u, \textbf{c}^v)/\sigma )$, and $p^u$ is rescaled within (0,1). The equivalence of Euclidean distance and cosine similarity is ensured by the $L_2$ normalization of Equation~(\ref{Equ: L2 Norm}).

\begin{table*}[t]
\centering
\setlength{\tabcolsep}{6pt}
\renewcommand{\arraystretch}{1.25}
\scalebox {0.66}
{
\begin{tabular}{cc|cccccc|cccccc|cccccc}
\hline
\multicolumn{2}{c|}{Datasets}                                                                                                                  & \multicolumn{6}{c|}{Tmall}                                                                                                                                                                                                                                               & \multicolumn{6}{c|}{Diginetica}                                                                                                                                                                                                                                          & \multicolumn{6}{c}{Retailrocket}                                                                                                                                                                                                                                         \\ \hline
\multicolumn{2}{c|}{Metrics}                                                                                                                   & \multicolumn{2}{c|}{Accuracy}                                                                        & \multicolumn{4}{c|}{Long-tail}                                                                                                                                    & \multicolumn{2}{c|}{Accuracy}                                                                        & \multicolumn{4}{c|}{Long-tail}                                                                                                                                    & \multicolumn{2}{c|}{Accuracy}                                                                        & \multicolumn{4}{c}{Long-tail}                                                                                                                                     \\ \hline
\multicolumn{1}{c|}{SBR Models}                                                                       & Methods                           & HR                                     & \multicolumn{1}{c|}{MRR}                                    & tHR                                    & tMRR                                   & tCov                                   & Tail                                   & HR                                     & \multicolumn{1}{c|}{MRR}                                    & tHR                                    & tMRR                                   & tCov                                   & Tail                                   & HR                                     & \multicolumn{1}{c|}{MRR}                                    & tHR                                    & tMRR                                   & tCov                                   & Tail                                   \\ \hline
\multicolumn{1}{c|}{}                                                                                 & \textit{base}                          & \underline{26.10}                            & \multicolumn{1}{c|}{\underline{14.67}}                            & \underline{25.98}                            & \underline{14.61}                            & 69.46                                  & 77.77                                  & \underline{50.15}                            & \multicolumn{1}{c|}{\underline{17.24}}                            & \underline{47.81}                            & \underline{16.56}                            & 90.71                                  & 68.70                                  & \underline{50.54}                            & \multicolumn{1}{c|}{\underline{26.34}}                            & \underline{49.66}                            & \underline{26.40}                            & 53.70                                  & 68.68                                  \\
\multicolumn{1}{c|}{}                                                                                 & \textit{+ TailNet}                     & 20.61                                  & \multicolumn{1}{c|}{9.91}                                   & 20.77                                  & 10.01                                  & 71.33                                  & \underline{78.01}                            & 45.39                                  & \multicolumn{1}{c|}{14.79}                                  & 42.68                                  & 14.89                                  & 91.23                                  & 68.21                                  & 47.00                                  & \multicolumn{1}{c|}{24.37}                                  & 46.21                                  & 24.21                                  & 51.56                                  & 63.76                                  \\
\multicolumn{1}{c|}{}                                                                                 & \textit{+ CSBR}                        & 25.43                                  & \multicolumn{1}{c|}{14.20}                                  & 25.46                                  & 14.28                                  & 69.15                                  & 77.58                                  & 49.86                                  & \multicolumn{1}{c|}{17.28}                                  & 47.80                                  & 16.48                                  & \underline{91.61}                            & 68.66                                  & 49.82                                  & \multicolumn{1}{c|}{25.96}                                  & 49.51                                  & 25.93                                  & 54.51                                  & 70.65                                  \\
\multicolumn{1}{c|}{}                                                                                 & \textit{+ LOAM}                        & 24.31                                  & \multicolumn{1}{c|}{13.80}                                  & 24.37                                  & 13.74                                  & 71.68                                  & 77.23                                  & 46.19                                  & \multicolumn{1}{c|}{15.28}                                  & 43.39                                  & 14.50                                  & 89.96                                  & \textbf{70.26}                         & 50.27                                  & \multicolumn{1}{c|}{26.13}                                  & 49.51                                  & 26.27                                  & \underline{55.67}                            & \underline{71.79}                            \\
\multicolumn{1}{c|}{}                                                                                 & \textit{+ LAP-SR}                      & 25.21                                  & \multicolumn{1}{c|}{14.13}                                  & 25.24                                  & 14.20                                  & \underline{72.11}                            & 77.61                                  & 49.87                                  & \multicolumn{1}{c|}{17.16}                                  & 47.69                                  & 16.37                                  & 91.32                                  & 68.55                                  & 49.59                                  & \multicolumn{1}{c|}{25.89}                                  & 48.78                                  & 25.93                                  & 55.32                                  & 71.41                                  \\
\multicolumn{1}{c|}{}                                                                                 &  \textbf{+ HID} &  \textbf{28.26} & \multicolumn{1}{c|}{ \textbf{15.84}} &  \textbf{28.35} &  \textbf{15.93} &  \textbf{73.65} &  \textbf{78.19} &  \textbf{50.39} & \multicolumn{1}{c|}{ \textbf{17.58}} &  \textbf{48.09} &  \textbf{17.28} &  \textbf{93.05} &  \underline{69.24}    &  \textbf{52.38} & \multicolumn{1}{c|}{ \textbf{27.99}} &  \textbf{52.09} &  \textbf{28.34} &  \textbf{56.02} &  \textbf{72.59} \\ \cline{2-20} 
\multicolumn{1}{c|}{\multirow{-7}{*}{\begin{tabular}[c]{@{}c@{}}STAMP\\ (Sequential)\end{tabular}}}   & \textit{p-value (\textless{})}         & 0.001                                  & \multicolumn{1}{c|}{0.001}                                  & 0.001                                  & 0.001                                  & 0.001                                  & 0.05                                   & 0.05                                   & \multicolumn{1}{c|}{0.05}                                   & 0.001                                   & 0.001                                  & 0.001                                  & 0.001                                  & 0.001                                  & \multicolumn{1}{c|}{0.001}                                  & 0.001                                  & 0.001                                  & 0.001                                  & 0.001                                  \\ \hline
\multicolumn{1}{c|}{}                                                                                 & \textit{base}                          & 19.69                                  & \multicolumn{1}{c|}{9.58}                                   & 19.53                                  & 9.57                                   & 49.60                                  & 71.80                                  & \underline{50.23}                            & \multicolumn{1}{c|}{\underline{16.96}}                            & \underline{47.49}                            & \underline{15.79}                            & 84.97                                  & 65.14                                  & 45.01                                  & \multicolumn{1}{c|}{\underline{24.33}}                            & 44.12                                  & 23.78                                  & 69.98                                  & 73.29                                  \\
\multicolumn{1}{c|}{}                                                                                 & \textit{+ TailNet}                     & 17.21                                  & \multicolumn{1}{c|}{8.25}                                   & 17.09                                  & 8.18                                   & 52.31                                  & 73.42                                  & 46.51                                  & \multicolumn{1}{c|}{15.30}                                  & 45.36                                  & 14.29                                  & 87.91                                  & 67.58                                  & 43.09                                  & \multicolumn{1}{c|}{22.98}                                  & 42.28                                  & 22.53                                  & 70.62                                  & 73.66                                  \\
\multicolumn{1}{c|}{}                                                                                 & \textit{+ CSBR}                        & \underline{19.90}                            & \multicolumn{1}{c|}{\underline{10.11}}                            & \underline{20.00}                            & \underline{10.27}                            & 53.22                                  & 78.52                                  & 49.92                                  & \multicolumn{1}{c|}{16.68}                                  & 47.01                                  & 15.43                                  & 88.74                                  & 68.24                                  & 43.39                                  & \multicolumn{1}{c|}{23.17}                                  & 43.41                                  & 22.71                                  & 70.99                                  & 74.21                                  \\
\multicolumn{1}{c|}{}                                                                                 & \textit{+ LOAM}                        & 18.40                                  & \multicolumn{1}{c|}{9.14}                                   & 18.65                                  & 9.31                                   & \underline{58.76}                            & \textbf{79.57}                         & 47.53                                  & \multicolumn{1}{c|}{15.65}                                  & 45.79                                  & 14.84                                  & \textbf{91.65}                         & \textbf{71.49}                         & \underline{45.32}                            & \multicolumn{1}{c|}{24.21}                                  & \underline{44.37}                            & \underline{23.89}                            & \underline{72.29}                            & \underline{75.19}                            \\
\multicolumn{1}{c|}{}                                                                                 & \textit{+ LAP-SR}                      & 19.41                                  & \multicolumn{1}{c|}{9.33}                                   & 19.37                                  & 9.29                                   & 56.32                                  & 76.12                                  & 49.91                                  & \multicolumn{1}{c|}{16.50}                                  & 46.98                                  & 15.31                                  & 90.21                                  & 68.03                                  & 44.59                                  & \multicolumn{1}{c|}{24.02}                                  & 43.67                                  & 23.61                                  & 71.45                                  & 74.03                                  \\
\multicolumn{1}{c|}{}                                                                                 &  \textbf{+ HID} &  \textbf{25.13} & \multicolumn{1}{c|}{ \textbf{13.95}} &  \textbf{25.21} &  \textbf{13.98} &  \textbf{63.21} &  \underline{77.92}    &  \textbf{52.23} & \multicolumn{1}{c|}{ \textbf{17.79}} &  \textbf{50.92} &  \textbf{16.83} &  \underline{90.73}    &  \underline{68.92}    &  \textbf{48.89} & \multicolumn{1}{c|}{ \textbf{26.43}} &  \textbf{47.91} &  \textbf{26.19} &  \textbf{73.21} &  \textbf{75.89} \\ \cline{2-20} 
\multicolumn{1}{c|}{\multirow{-7}{*}{\begin{tabular}[c]{@{}c@{}}GRU4Rec\\ (Sequential)\end{tabular}}} & \textit{p-value (\textless{})}         & 0.001                                  & \multicolumn{1}{c|}{0.001}                                  & 0.001                                  & 0.001                                  & 0.001                                  & -                                      & 0.001                                  & \multicolumn{1}{c|}{0.001}                                  & 0.001                                  & 0.001                                  & -                                      & -                                      & 0.001                                  & \multicolumn{1}{c|}{0.001}                                  & 0.001                                  & 0.001                                  & 0.001                                  & 0.005                                  \\ \hline
\multicolumn{1}{c|}{}                                                                                 & \textit{base}                          & \underline{27.45}                            & \multicolumn{1}{c|}{\underline{14.27}}                            & \underline{27.12}                            & \underline{14.32}                            & 53.60                                  & 77.65                                  & \underline{51.47}                            & \multicolumn{1}{c|}{\underline{17.95}}                            & 49.04                                  & 17.01                                  & 94.16                                  & 68.85                                  & \underline{50.55}                            & \multicolumn{1}{c|}{\underline{26.88}}                            & \underline{49.16}                            & \underline{26.24}                            & 53.96                                  & 69.94                                  \\
\multicolumn{1}{c|}{}                                                                                 & \textit{+ TailNet}                     & 25.79                                  & \multicolumn{1}{c|}{13.05}                                  & 25.81                                  & 13.39                                  & 64.01                                  & 76.33                                  & 49.86                                  & \multicolumn{1}{c|}{17.42}                                  & 48.21                                  & 16.98                                  & 88.97                                  & 65.77                                  & 47.87                                  & \multicolumn{1}{c|}{25.14}                                  & 47.11                                  & 24.78                                  & 54.13                                  & 71.47                                  \\
\multicolumn{1}{c|}{}                                                                                 & \textit{+ CSBR}                        & 26.98                                  & \multicolumn{1}{c|}{13.83}                                  & 26.89                                  & 13.90                                  & 53.00                                  & 77.61                                  & 51.22                                  & \multicolumn{1}{c|}{17.89}                                  & \underline{49.16}                            & \underline{17.02}                            & 93.89                                  & 68.94                                  & 49.93                                  & \multicolumn{1}{c|}{26.55}                                  & 48.76                                  & 25.79                                  & 55.32                                  & 71.65                                  \\
\multicolumn{1}{c|}{}                                                                                 & \textit{+ LOAM}                        & 26.33                                  & \multicolumn{1}{c|}{13.52}                                  & 26.56                                  & 13.75                                  & \textbf{69.95}                         & \underline{77.23}                            & 49.27                                  & \multicolumn{1}{c|}{17.19}                                  & 48.03                                  & 16.70                                  & \underline{95.92}                            & \textbf{72.11}                         & 50.29                                  & \multicolumn{1}{c|}{26.81}                                  & 49.02                                  & 26.20                                  & \textbf{56.16}                         & \textbf{73.97}                         \\
\multicolumn{1}{c|}{}                                                                                 & \textit{+ LAP-SR}                      & 26.76                                  & \multicolumn{1}{c|}{13.95}                                  & 26.89                                  & 14.05                                  & 61.35                                  & 75.38                                  & 51.04                                  & \multicolumn{1}{c|}{17.84}                                  & 48.86                                  & 16.92                                  & 95.22                                  & 71.94                                  & 50.32                                  & \multicolumn{1}{c|}{26.37}                                  & 48.76                                  & 26.02                                  & 54.99                                  & 71.59                                  \\
\multicolumn{1}{c|}{}                                                                                 &  \textbf{+ HID} &  \textbf{28.38} & \multicolumn{1}{c|}{ \textbf{14.66}} &  \textbf{28.13} &  \textbf{14.50} &  \underline{66.40}    &  \textbf{78.12} &  \textbf{52.09} & \multicolumn{1}{c|}{ \textbf{18.26}} &  \textbf{49.79} &  \textbf{17.25} &  \textbf{96.22} &  \underline{70.05}    &  \textbf{53.45} & \multicolumn{1}{c|}{ \textbf{29.47}} &  \textbf{52.61} &  \textbf{29.51} &  \underline{55.75}    &  \underline{73.54}    \\ \cline{2-20} 
\multicolumn{1}{c|}{\multirow{-7}{*}{\begin{tabular}[c]{@{}c@{}}SRGNN\\ (Graphic)\end{tabular}}}      & \textit{p-value (\textless{})}         & 0.001                                  & \multicolumn{1}{c|}{0.001}                                  & 0.001                                  & 0.05                                   & -                                      & 0.01                                   & 0.001                                  & \multicolumn{1}{c|}{0.001}                                  & 0.001                                  & 0.05                                   & 0.005                                    & -                                      & 0.001                                  & \multicolumn{1}{c|}{0.001}                                  & 0.001                                  & 0.001                                  & -                                      & -                                      \\ \hline
\multicolumn{1}{c|}{}                                                                                 & \textit{base}                          & \underline{32.42}                            & \multicolumn{1}{c|}{\underline{13.98}}                            & \underline{32.35}                            & \underline{13.94}                            & 81.99                                  & 77.49                                  & \underline{53.84}                            & \multicolumn{1}{c|}{\underline{18.87}}                            & \underline{51.55}                            & \underline{18.04}                            & 91.43                                  & 45.82                                  & \underline{54.97}                            & \multicolumn{1}{c|}{\underline{28.47}}                            & \underline{54.61}                            & \underline{28.13}                            & 72.54                                  & 72.13                                  \\
\multicolumn{1}{c|}{}                                                                                 & \textit{+ TailNet}                     & 29.91                                  & \multicolumn{1}{c|}{12.50}                                  & 29.70                                  & 12.41                                  & 83.76                                  & 77.91                                  & 47.47                                  & \multicolumn{1}{c|}{16.78}                                  & 45.59                                  & 15.90                                  & 92.13                                  & 46.01                                  & 53.19                                  & \multicolumn{1}{c|}{27.57}                                  & 52.87                                  & 27.33                                  & 73.85                                  & 72.73                                  \\
\multicolumn{1}{c|}{}                                                                                 & \textit{+ CSBR}                        & 29.60                                  & \multicolumn{1}{c|}{14.07}                                  & 29.37                                  & 13.86                                  & 82.80                                  & 78.21                                  & 52.24                                  & \multicolumn{1}{c|}{18.06}                                  & 50.13                                  & 17.35                                  & \underline{93.81}                            & \underline{46.32}                            & 52.26                                  & \multicolumn{1}{c|}{26.90}                                  & 51.99                                  & 26.45                                  & 73.23                                  & 72.59                                  \\
\multicolumn{1}{c|}{}                                                                                 & \textit{+ LOAM}                        & 30.96                                  & \multicolumn{1}{c|}{13.54}                                  & 30.79                                  & 13.47                                  & \textbf{84.97}                         & \underline{78.11}                            & 52.31                                  & \multicolumn{1}{c|}{17.42}                                  & 50.19                                  & 16.77                                  & 93.01                                  & 46.13                                  & 53.78                                  & \multicolumn{1}{c|}{27.81}                                  & 53.47                                  & 27.56                                  & \textbf{75.11}                         & \underline{74.47}                            \\
\multicolumn{1}{c|}{}                                                                                 & \textit{+ LAP-SR}                      & 32.11                                  & \multicolumn{1}{c|}{13.67}                                  & 32.05                                  & 13.63                                  & 82.03                                  & 77.89                                  & 53.19                                  & \multicolumn{1}{c|}{18.20}                                  & 50.89                                  & 17.51                                  & 92.44                                  & 45.91                                  & 53.26                                  & \multicolumn{1}{c|}{27.40}                                  & 52.96                                  & 27.02                                  & 74.02                                  & 73.97                                  \\
\multicolumn{1}{c|}{}                                                                                 &  \textbf{+ HID} &  \textbf{33.53} & \multicolumn{1}{c|}{ \textbf{14.43}} &  \textbf{33.31} &  \textbf{14.37} &  \underline{83.25}    &  \textbf{78.70} &  \textbf{54.22} & \multicolumn{1}{c|}{ \textbf{19.18}} &  \textbf{51.83} &  \textbf{18.37} &  \textbf{94.21} &  \textbf{46.67} &  \textbf{55.37} & \multicolumn{1}{c|}{ \textbf{28.82}} &  \textbf{54.99} &  \textbf{28.59} &  \underline{74.74}    &  \textbf{74.89} \\ \cline{2-20} 
\multicolumn{1}{c|}{\multirow{-7}{*}{\begin{tabular}[c]{@{}c@{}}GCEGNN\\ (Graphic)\end{tabular}}}     & \textit{p-value (\textless{})}         & 0.001                                  & \multicolumn{1}{c|}{0.001}                                  & 0.001                                  & 0.001                                  & -                                      & 0.001                                  & 0.005                                  & \multicolumn{1}{c|}{0.001}                                  & 0.005                                  & 0.05                                   & 0.001                                  & 0.05                                   & 0.001                                  & \multicolumn{1}{c|}{0.001}                                  & 0.05                                   & 0.005                                  & -                                      & 0.01                                   \\ \hline
\end{tabular}}
\caption{The accuracy and long-tail performance (K=20) of SBR models with long-tail methods over three datasets. Bold labeled scores indicate the best results for each dataset under certain baseline and underlined scores represent second-best results. The p-value is calculated through two-sided t-test.}
\label{Table: Overall}
\end{table*}

\begin{table*}[ht]
\setlength{\tabcolsep}{6pt}
\renewcommand{\arraystretch}{1.2}
\scalebox {0.66}{
\begin{tabular}{cc|cccccc|cccccc|cccccc}
\hline
\multicolumn{2}{c|}{Datasets}                                              & \multicolumn{6}{c|}{Tmall}                                                                                                                                                                                                                                               & \multicolumn{6}{c|}{Diginetica}                                                                                                                                                                                                                                          & \multicolumn{6}{c}{Retailrocket}                                                                                                                                                                                                                                         \\ \hline
\multicolumn{1}{c|}{SBR Model}               & Comparisons                 & HR                                     & \multicolumn{1}{c|}{MRR}                                    & tHR                                    & tMRR                                   & tCov                                   & Tail                                   & HR                                     & \multicolumn{1}{c|}{MRR}                                    & tHR                                    & tMRR                                   & tCov                                   & Tail                                   & HR                                     & \multicolumn{1}{c|}{MRR}                                    & tHR                                    & tMRR                                   & tCov                                   & Tail                                   \\ \hline
\multicolumn{1}{c|}{}                        &  HID &  \textbf{28.26} & \multicolumn{1}{c|}{ \textbf{15.84}} &  \textbf{28.35} &  \textbf{15.93} &  \textbf{73.65} &  \textbf{78.19} &  \textbf{50.39} & \multicolumn{1}{c|}{ \textbf{17.38}} &  \textbf{48.09} &  \textbf{17.28} &  \textbf{93.05} &  \textbf{69.24} &  \textbf{52.38} & \multicolumn{1}{c|}{ \textbf{27.99}} &  \textbf{52.09} &  \textbf{28.34} &  \textbf{56.02} &  \textbf{72.59} \\
\multicolumn{1}{c|}{}                        & HID w/o HI                  & 27.43                                  & \multicolumn{1}{c|}{15.24}                                  & 27.56                                  & 15.40                                  & 69.29                                  & 77.98                                  & 50.17                                  & \multicolumn{1}{c|}{17.24}                                  & 47.96                                  & 17.19                                  & 91.96                                  & 68.83                                  & 51.75                                  & \multicolumn{1}{c|}{27.37}                                  & 51.51                                  & 27.82                                  & 55.31                                  & 71.80                                  \\
\multicolumn{1}{c|}{\multirow{-3}{*}{STAMP}} & HID w/o FC                  & 26.77                                  & \multicolumn{1}{c|}{14.86}                                  & 26.86                                  & 15.09                                  & 70.20                                  & 77.94                                  & 49.76                                  & \multicolumn{1}{c|}{17.24}                                  & 47.52                                  & 16.31                                  & 92.15                                  & 68.91                                  & 50.89                                  & \multicolumn{1}{c|}{26.51}                                  & 50.60                                  & 26.71                                  & 55.67                                  & 72.16                                  \\ \hline
\multicolumn{1}{c|}{}                        &  HID &  \textbf{28.38} & \multicolumn{1}{c|}{ \textbf{14.66}} &  \textbf{28.13} &  \textbf{14.50} &  \textbf{66.40} &  \textbf{78.12} &  \textbf{52.09} & \multicolumn{1}{c|}{ \textbf{18.26}} &  \textbf{49.79} &  \textbf{17.25} &  \textbf{96.02} &  \textbf{70.05} &  \textbf{53.45} & \multicolumn{1}{c|}{ \textbf{29.47}} &  \textbf{52.61} &  \textbf{29.51} &  \textbf{55.75} &  \textbf{73.54} \\
\multicolumn{1}{c|}{}                        & HID w/o HI                  & 27.48                                  & \multicolumn{1}{c|}{14.34}                                  & 27.31                                  & 14.36                                  & 61.00                                  & 77.12                                  & 51.96                                  & \multicolumn{1}{c|}{18.01}                                  & 49.46                                  & 17.03                                  & 92.94                                  & 68.57                                  & 53.10                                  & \multicolumn{1}{c|}{29.18}                                  & 52.27                                  & 29.21                                  & 54.01                                  & 72.77                                  \\
\multicolumn{1}{c|}{\multirow{-3}{*}{SRGNN}} & HID w/o FC                  & 27.36                                  & \multicolumn{1}{c|}{14.33}                                  & 27.23                                  & 14.27                                  & 62.92                                  & 77.49                                  & 51.16                                  & \multicolumn{1}{c|}{17.40}                                  & 48.90                                  & 16.41                                  & 93.56                                  & 69.10                                  & 52.80                                  & \multicolumn{1}{c|}{28.79}                                  & 51.98                                  & 28.83                                  & 55.11                                  & 73.03                                  \\ \hline
\end{tabular}}
\caption{Ablation study on Tmall, Diginetica and Retailrocket.}
\label{Table: Ablation}
\end{table*}

\subsubsection{Multi-task Learning.}
\label{Sect: Multi-task Learning}
To incorporate HID into traditional SBR models, we introduce a multi-task learning loss to combine the learning of ICLoss with the typically used cross-entorpy loss. Specifically, a hyper-parameter $\epsilon$ is introduced to control the scale of ICLoss. The total loss can be expressed as: $\mathcal{L} = \mathcal{L}_p + \epsilon\mathcal{L}_{c}$. Besides, the time complexity analysis of HID is provided in \textbf{Appendix}~C.

\section{Experiments}
\label{Sect: Experiments}
\textbf{Datasets}. We evaluate our proposed HID with the three real-world datasets, namely  $\mathit{Tmall}$\footnote{https://tianchi.aliyun.com/dataset/dataDetail?dataId=42}, $\mathit{RetailRocket}$\footnote{https://www.kaggle.com/retailrocket/ecommerce-dataset}, $\mathit{Diginetica}$\footnote{https://competitions.codalab.org/competitions/11161}. 
\textit{Tmall} is from the IJCAI-15 competition and consists of shopping logs of unnamed users on the Tmall online shopping platform. \textit{RetailRocket} RetailRocketis released by an e-commerce corporation for the Kaggle competition and contains users’ browsing activity. \textit{Diginetica} comes from CIKM Cup 2016.

\textbf{Base Models and Competitors}. To demonstrate the effectiveness of our proposed HID, we select some well-known SBR models from both sequential approaches (GRU4Rec~\cite{Hidasi16gru4rec}, STAMP~\cite{liu18Stamp}) and graphic approaches (SR-GNN~\cite{Wu19Srgnn}, GCE-GNN~\cite{Wang20Gcegnn}) as the base SBR models. Apart from the above base SBR models, we also introduce TailNet~\cite{Liu2020TailNet}, CSBR~\cite{Chen2023CSBR}, LOAM~\cite{Yang2023LOAM}, LAP-SR~\cite{Peng2024LAPSR} as the plug-and-play long-tail competitors.

\textbf{Metrics}. To evaluate the recommendation accuracy and long-tail performance, we employ three widely used accuracy metrics, including the HR@K, and MRR@K. Following previous works on long-tail issue~\cite{Himan19Managing, Liu2020TailNet, Yang2023LOAM}, we introduce some well-known long-tail metrics, including tHR@K, tMRR@K, tCov@K, and Tail@K.

More details on the preprocessing process, baselines, metrics and implementation details are given in \textbf{Appendix}~D.

\subsection{Ablation Study}
To investigate our proposed method, we construct two variants of our proposed method which are the HID w/o HI (i.e., Hybrid Intent) where the hybrid intent is substituted with the commonly used intent definition based on the last few items (3 in this experiment, and average pooling is adopted to aggregate them) of each session~\cite{Zhang23Efficiently}, and the HID w/o FC (i.e., Flexible Coefficient) where the flexible coefficient $\sigma$ is dropped. Experiments are demonstrated in Table~\ref{Table: Ablation}. Overall, both HID w/o HI and HID w/o FC exhibit performance degradation compared to HID across the two SBR models and datasets. Removing HI impacts diversity more, while removing FC affects accuracy more, consistent with our prior analysis. Furthermore, the hybrid intents have greater impact on Tmall than on Diginetica/Retailrocket, as its longer sessions exhibit more frequent intent shifts, making target intent modeling crucial.

\subsection{Overall Performance}
\label{Sect: Experiments-Overall Performance}
Refer to results in Table~\ref{Table: Overall}, we draw following conclusions:

\textbf{For Previous Work}. The results indicate that almost all existing long-tail approaches improve long-tail performance with the \textit{sacrifice of accuracy} compared with base SBR models. This trade-off arises from their neglect of the substantial amount of session-irrelevant items, which introduces noise into the recommendations when prioritizing tail items.

\textbf{For Our Proposed HID}. Compared with previous appraoches, SBR models with HID demonstrate improvements in both accuracy and long-tail performance. This improvement arises from two aspects: (i) The representative hybrid intent endows HID with the capability to perceive users' high-level intents, providing a solid foundation for the effectiveness of the overall framework; (ii) The intent constraint loss effectively emphasizes tail items within the target intent while driving session representations away from noise distributions, thus achieving accurate long-tail SBR.

\begin{figure}[t]
    \centering
    \subfloat[Tmall]{\includegraphics[width=0.314\linewidth]{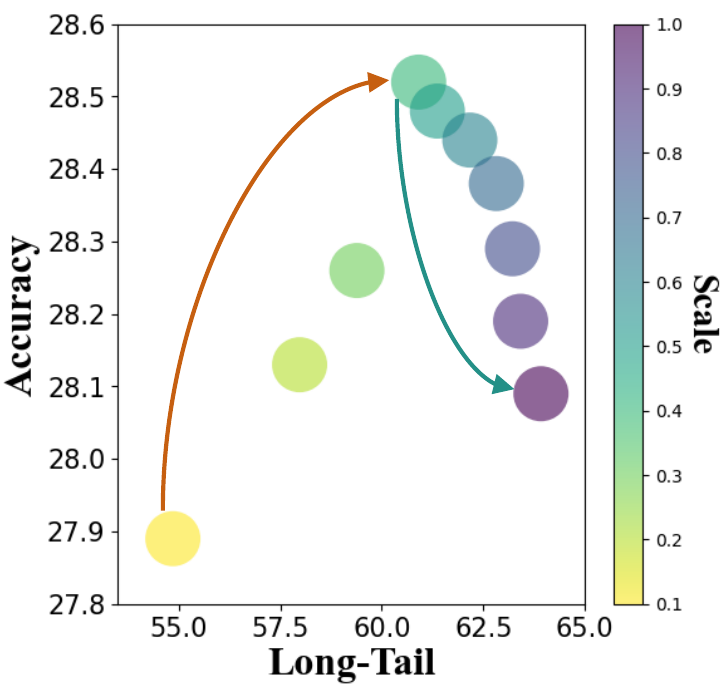}}
    \hfill
    \subfloat[Diginetica]{\includegraphics[width=0.318\linewidth]{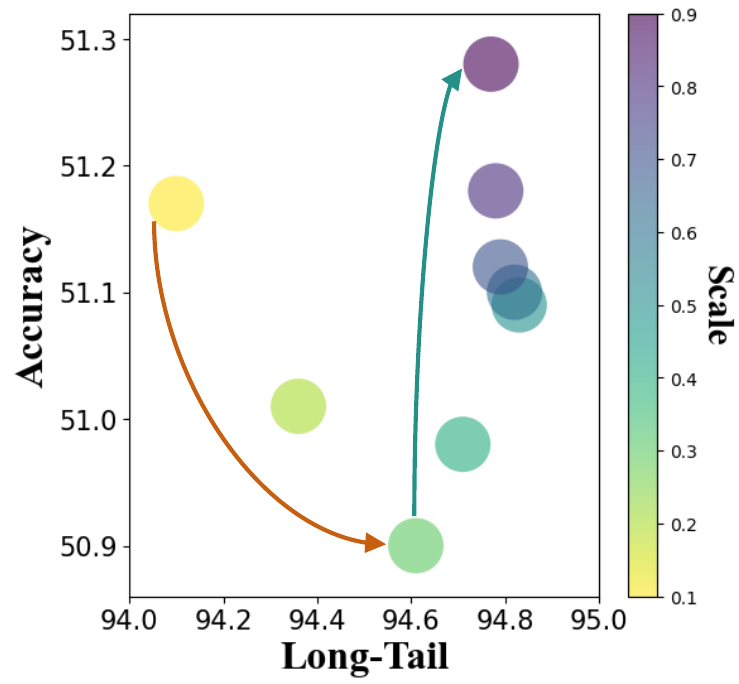}}
    \hfill
    \subfloat[Retailrocket]{\includegraphics[width=0.318\linewidth]{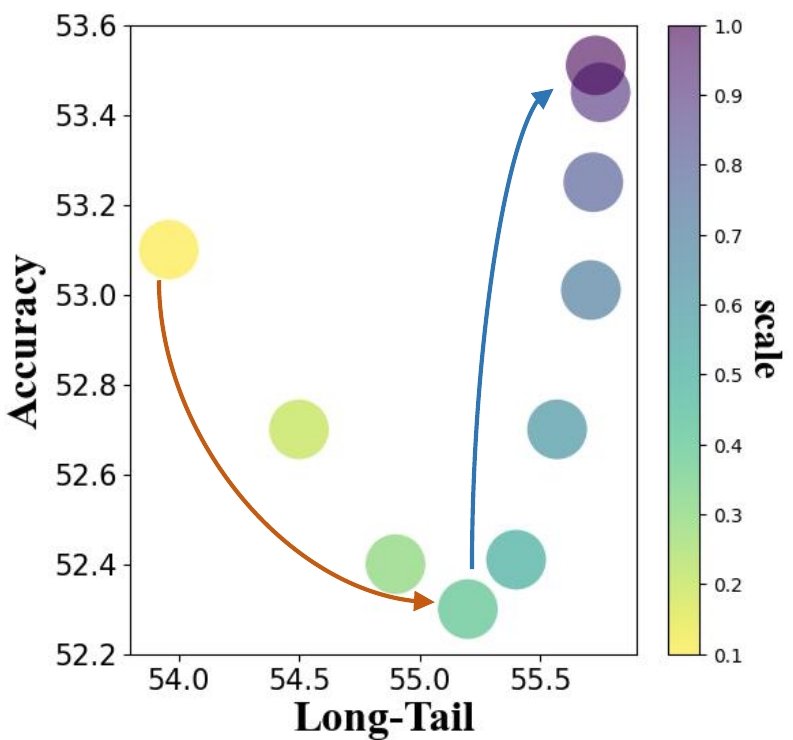}}
    \caption{The changes in accuracy (HR@20) and long-tail (tCov@20) metrics with the increase of scale $\epsilon$. The model is SRGNN+HID.}
    \label{Fig: Scale}
\end{figure}

\begin{figure}[t]
    \centering
    \subfloat[Tmall]{\includegraphics[width=0.314\linewidth]{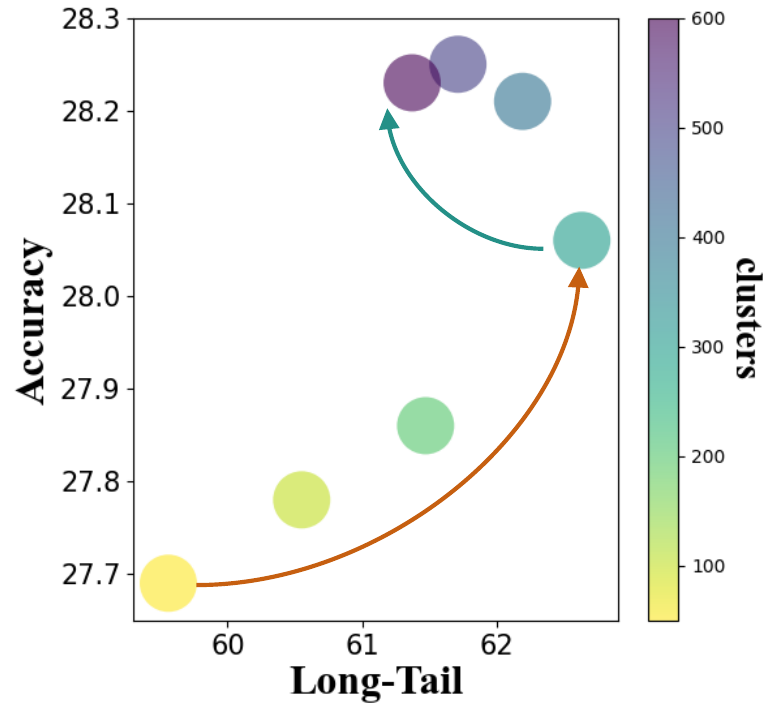}}
    \hfill
    \subfloat[Diginetica]{\includegraphics[width=0.314\linewidth]{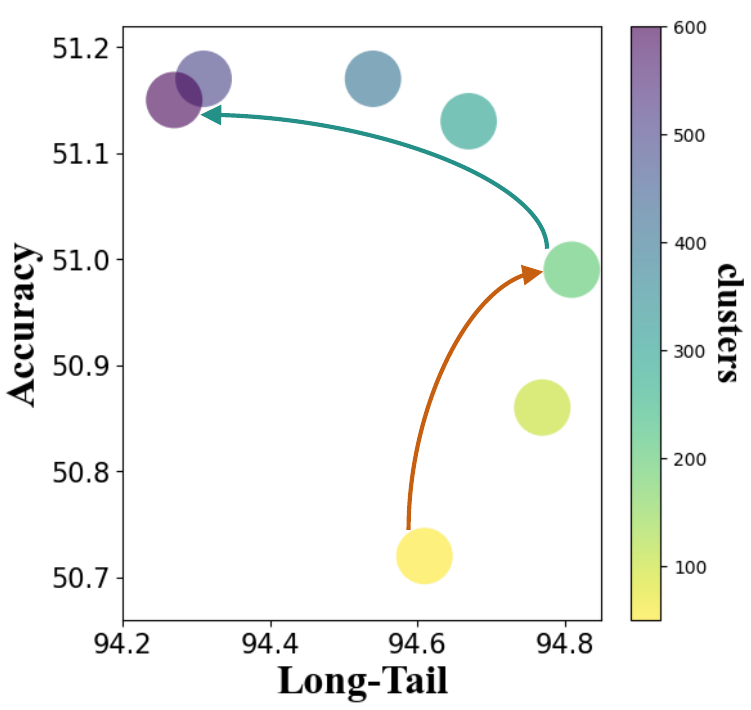}}
    \hfill
    \subfloat[Retailrocket]{\includegraphics[width=0.318\linewidth]{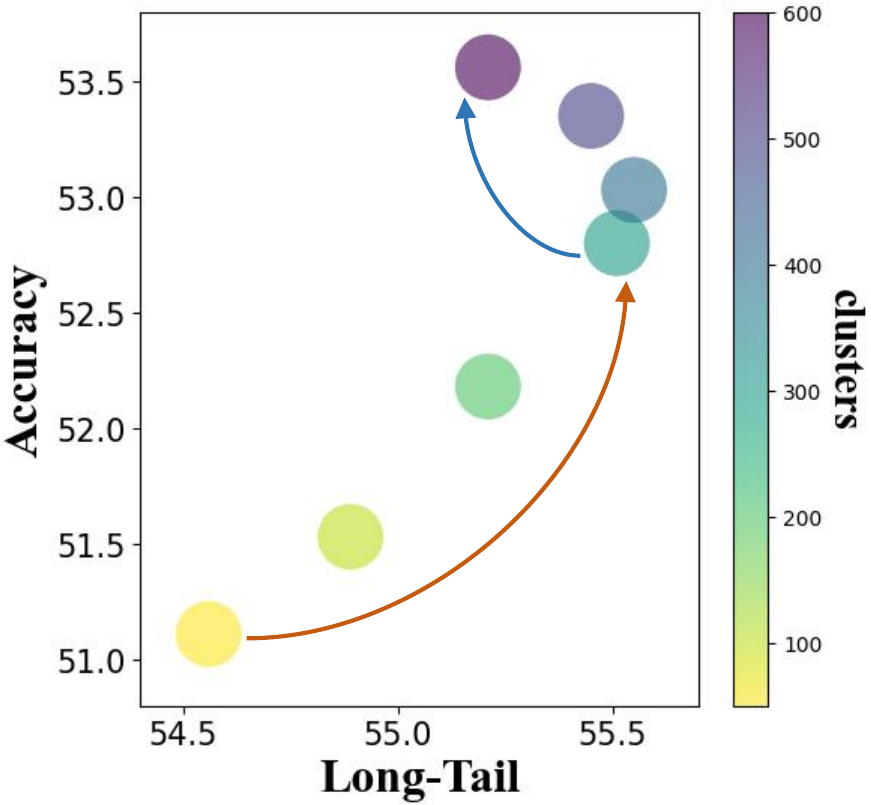}}
    \caption{The changes in accuracy (HR@20) and long-tail (tCov@20) metrics with the increase of clusters $n$. The model is SRGNN+HID.}
    \label{Fig: Cluster}
\end{figure}

\subsection{Hyperparameter Exploration}
\subsubsection{Balance between ICLoss and CE Loss.}
We systematically study the balance between cross-entropy loss and ICLoss by tuning the scaling parameter $n$ from 0.1 to 0.9. As shown in Figure~\ref{Fig: Scale}, the Tmall dataset demonstrates distinct behavior: both accuracy (HR@20) and long-tail performance (tCov@20) improve as clusters $n$ increases from 0 to 0.4, beyond which accuracy declines while long-tail performance continues to improve, establishing $n$=0.4 as the optimal trade-off point. In contrast, Diginetica and Retailrocket exhibit different patterns - their long-tail performance initially improves then stabilizes with increasing $n$, while accuracy shows non-monotonic variations. Therefore, on SRGNN, for the Tmall dataset, increasing the weight of ICLoss in the range from 0.1 to 0.4 can further improve both accuracy and long-tail performance. For Diginetica and Retailrocket, the range is from 0.3 to 0.9 and 0.4 to 0.9.

\subsubsection{Number of Hybrid Intents.}
In this section, we investigate the impact of cluster numbers (i.e., number of hybrid intents) $n$ of spectral clustering on the recommendation accuracy and long-tail performance. As shown in Figure~\ref{Fig: Cluster}, on both Tmall and Diginetica, we observe the same trend that as the number of hybrid intents increasing, the accuracy increases initially and then stabilizes while the long-tail performance exhibits a peak-shaped pattern, reaching its maximum when the number of clusters is 4 for Tmall and Retailrocket, and 3 for Diginetica. This indicates that when the number of hybrid intents increases (i.e., each intent contains fewer items), more items are classified as noise intents. As a result, HID is able to exclude more noise items from the recommendation list. However, for the diversity metric, there is less items belonging to the target intent, which leads to fewer long-tail items being considered by HID, causing a decline in long-tail performance.

\subsection{Replacing Attribute with Semantic Clusters}
We conduct additional experiments to investigate the performance of HID without attributes in the \textbf{Appendix D}.

\section{Conclusion}
This paper addresses the challenge of balancing long-tail performance and recommendation accuracy in traditional SBR methods by proposing a Hybrid Intent-based Dual Constraint Framework (HID), transforming the typical "see-saw" into the "win-win". We introduce two novel constraints targeting both long-tail performance and recommendation accuracy, enforced through a hybrid intent learning process that captures both the attributes of items and actions of anonymous users. Additionally, we propose the intent constraint loss (ICLoss), which guides session representation learning and integrates seamlessly with existing SBR models. Extensive experiments on multiple baselines and datasets validate effectiveness of HID, proving that it can improve both accuracy and long-tail performance for SBR.

\bibliography{aaai2026}

\appendix
\section*{A~~~Proof of Theorem 1 \\(Optimizing Equivalence)}
\label{Apdx: Proof of Lemma 1}
\textbf{Theorem 1} (Optimizing Equivalence). \textit{The Equation~(6) (in the main paper file) with time complexity of $O(N \times d)$ can be approximated to an equation with time complexity of $O(d)$:}
\begin{align}
    &\min~\mathcal{L}_l =\text{Var}_{v_i \in \mathcal{C}^u}[d(\textbf{S}^u, \textbf{v}_i)]~~\sim ~~\min~d(\textbf{S}^u, \textbf{c}^u).
\label{Equ: apdx CoL2}
\end{align}

\textit{Proof}. The Theorem 1 can be proved as follows: 
\begin{align}
\begin{split}
    \mathcal{L}_l = &~\text{Var}_{\textbf{v}_i \in \mathcal{C}^u}[d(\textbf{S}^u, \textbf{v}_i)] \\
    = &~\underset{\textbf{v}_i\in \mathcal{C}^u}{\sum}\frac{\|\textbf{S}^u-\textbf{v}_i\|^2}{{|\mathcal{C}^u|}}-\left(\underset{\textbf{v}_i\in \mathcal{C}^u}{\sum}\frac{\|\textbf{S}^u-\textbf{v}_i\|}{{|\mathcal{C}^u|}}\right)^2.
\end{split}
\label{Equ: variance1}
\end{align}

Then, we calculate the gradient of the $\mathcal{L}_l$:
\begin{align}
\begin{split}
    \nabla\mathcal{L}_l=~2(\textbf{S}^u-\underset{\textbf{v}_i\in \mathcal{C}^u}{\sum}\frac{\textbf{v}_i}{|\mathcal{C}^u|})-2\left(\underset{\textbf{v}_i\in \mathcal{C}^u}{\sum}\frac{\|\textbf{S}^u-\textbf{v}_i\|}{{|\mathcal{C}^u|}}\right) \times
     \\ \left(\underset{\textbf{v}_i\in \mathcal{C}^u}{\sum}\frac{\textbf{S}^u-\textbf{v}_i}{{|\mathcal{C}^u|}\cdot\|\textbf{S}^u-\textbf{v}_i\|}\right).
\end{split}
\label{Equ: variance2}
\end{align}

After setting it to zero, we obtain the expression for $\textbf{S}^u$:
\begin{align}
\begin{split}
    \textbf{S}^u = \underset{\textbf{v}_i\in \mathcal{C}^u}{\sum}\frac{\textbf{v}_i}{|\mathcal{C}^u|}
+\left(\underset{\textbf{v}_i\in \mathcal{C}^u}{\sum}\frac{\|\textbf{S}^u-\textbf{v}_i\|}{{|\mathcal{C}^u|}}\right)\times \\ \left(\underset{\textbf{v}_i\in \mathcal{C}^u}{\sum}\frac{\textbf{S}^u-\textbf{v}_i}{{|\mathcal{C}^u|}\cdot\|\textbf{S}^u-\textbf{v}_i\|}\right).
\end{split}
\label{Equ: variance3}
\end{align}

Considering that the second term $\underset{\textbf{v}_i\in \mathcal{C}^u}{\sum}\frac{\textbf{S}^u-\textbf{v}_i}{{|\mathcal{C}^u|}\cdot\|\textbf{S}^u-\textbf{v}_i\|} \sim 0$ when the sum of the unit vectors pointing from $\textbf{S}^u$ to each point $\textbf{v}_i$ is about 0, which can be satisfied when $\textbf{v}_i$ exhibit an approximately symmetric distribution around $\textbf{S}^u$, indicating that $\textbf{S}^u$ is the centorid of all $\textbf{v}_i$. Observing the Equation~(18), when the second term is approximated to 0, $S^u$ will be approximately equal to the first term $\underset{\textbf{v}_i\in \mathcal{C}^u}{\sum}\frac{\textbf{v}_i}{|\mathcal{C}^u|}$, which is exactly the centorid of all $\textbf{v}_i$. 

In that case, $\textbf{S}^u = \underset{\textbf{v}_i\in \mathcal{C}^u}{\sum}\frac{\textbf{v}_i}{|\mathcal{C}^u|}$ is an approximate solution of optimizing Equation~(16). Referring to Equation~(2), the target intent embedding is aquired through the average pooling: $\textbf{c}^u = \underset{v_i \in \mathcal{C}^u}{\sum}\frac{\textbf{v}_i}{|\mathcal{C}^u|}$. Therefore, minimizing $d(\textbf{S}^u, \textbf{c}^u)$ can be approximately equal to minimizing Equation~(6).

\section*{B~~~Proof of Theorem 2 \\(Triplet Loss Approximation)}
\label{Apdx: Proof of Lemma 2}
\textbf{Theorem 2} (Triplet Loss Approximation). \textit{The optimization of the objective function in Equation~(9) is approximately proportional to optimize a (N-1)-triplet loss with a fixed margin of 2}:
\begin{align}
    \begin{split}
    &\mathcal{L}_c \propto\sum_{S^u \in \mathcal{B}} \sum_{c^{v} \in \mathcal{\hat{C}}^u}\left(\|\textbf{S}^u-\textbf{c}^u\|^{2}-\|\textbf{S}^u-\textbf{c}^v\|^{2}+\textbf{2}\right),
    \end{split}
\label{Equ: (N-1)TupletLoss)}
\end{align}

\textit{Proof}. Since the $\textbf{S}^u$ and $\textbf{c}^u$ have been $L_2$ normalized before sending to ICLoss, minimizing $d(\textbf{S}^u,\textbf{c}^u)$ is equivalent to maximizing $\textbf{S}^u\cdot \textbf{c}^u$. Then, the Theorem 2 can be proved as follows: 
\begin{equation}
    \begin{split}
    \mathcal{L}_c &= - \sum_{S^u \in \mathcal{B}}\text{log}\frac{\text{exp}(\textbf{S}^u\cdot \textbf{c}^u)}{\text{exp}(\textbf{S}^u\cdot \textbf{c}^u)+\sum_{c^{v} \in \mathcal{\hat{C}}^u}\text{exp}(\textbf{S}^u\cdot \textbf{c}^v)},\\ 
    &=\sum_{S^u \in \mathcal{B}}\text{log}[1+\sum_{c^{v} \in \mathcal{\hat{C}}^u}\text{exp}(\textbf{S}^u\cdot \textbf{c}^v - \textbf{S}^u\cdot \textbf{c}^u)],\\ \nonumber
    &\simeq \sum_{S^u \in \mathcal{B}}\underset{c^{v} \in \mathcal{\hat{C}}^u}{\sum}\text{exp}(\textbf{S}^u\cdot \textbf{c}^v - \textbf{S}^u\cdot \textbf{c}^u),\\ \nonumber
    &\simeq \sum_{S^u \in \mathcal{B}}\underset{c^{v} \in \mathcal{\hat{C}}^u}{\sum}(\textbf{S}^u\cdot \textbf{c}^v - \textbf{S}^u\cdot \textbf{c}^u+1),\\
    &\propto \sum_{S^u \in \mathcal{B}}\underset{c^{v} \in \mathcal{\hat{C}}^u}{\sum}(||\textbf{S}^u-\textbf{c}^u||^2-||\textbf{S}^u-\textbf{c}^v||^2+2).
    \end{split}
\label{Equ: Lemma2Proof}
\end{equation}

\section*{C~~~Time Complexity Analysis}
\label{Apdx: Complexity}
The main components of HID are the hybrid intent learning module and the intent constraint loss. In the hybrid intent learning module, since the item attributes and the connections between attributes across all sessions can be pre-obtained from the dataset, we construct the preliminary intent graph for each dataset and store the results of spectral clustering in advance. In that case, the complexity of this module arises solely from the average pooling used to obtain the hybrid intent representation, which is $O(md)$. For the intent constraint loss, for each batch, the complexity of the \textit{Constraint for long-tail} is $O(Bd)$ where B is the batch size, the complexity of the \textit{Constraint for Accuracy} is $O(B Kd)$ where $K$ is the average number of noise intents for sessions. Therefore, the complexity of the intent constraint loss for each batch is $O(B(d+Kd)) = O(BKd)$ since $K$ is typically larger than $d$.

\section*{D~~~Experimental Details}
\label{Apdx: Experimental Details}

\label{Apdx: Experimental Details}
\subsection*{D.1~~~Preprocess of the Datasets.}
Following \cite{Xia21Cotrec}, we conduct preprocessing steps over each dataset. Specifically, sessions with a length of 1 and items that appeared fewer than 5 times are excluded. Similar to \cite{Wang20Gcegnn}, we set the sessions of last week (i.e., latest data) as the test data, and the remaining historical data for training. Additionally, we use a session splitting preprocess method to augment session $S=\{s_{1}, s_{2},..., s_{n}\}$ in these datasets, and generate multiple sessions with corresponding labels $([s_{1}, s_{2}]; s_{3}),([s_{1}, s_{2}, s_{3}]; s_{4}), ...,([s_{1}, s_{2},..., s_{n-1}];s_{n})$. The statistics of the datasets are presented in Table~1.

\subsection*{D.2~~~Computation Resources}
\label{Apdx: Experimental Details-Computation Resources}
 The experiments are run on Linux with Intel(R) Xeon(R) Gold 6342 CPU with max CPU speed of 2.80GHz. We implement all the algorithms in this paper using PyTorch. All algorithms are run with a single Nvidia GeForce RTX 3090 GPU.

\subsection*{D.3~~~SBR Models and Comparison Approaches.}
From the perspective of data modeling, we select some well-known SBR models from both sequentail and graphic approaches as the base SBR models:
\begin{itemize}
\item \textbf{STAMP}~\cite{liu18Stamp} explores the capability of attention layers on session-based recommendation instead of RNNs. It optimizes the attention mechanism of previous work by emphasizing the user's short-term memory.
\item  \textbf{GRU4Rec}~\cite{Hidasi16gru4rec} is An RNN based deep learning model for session based recommendation, which utilizes session-parallel mini-batch training process and also
employs ranking-based loss functions during the training.
\item \textbf{SR-GNN}~\cite{Wu19Srgnn} employs GNNs to learn item embeddings and fuse the item-level information to get the session representation by leveraging the soft-attention mechanism.
\item \textbf{GCE-GNN}~\cite{Wang20Gcegnn} constructs two types of graphs to capture the global and local information from input sessions and combine them to enhance the feature presentations of items.
\end{itemize}

\begin{table}[t]
\centering
\begin{tabular}{llll}
\hline
Dataset           & Tmall   & RetailRocket & Diginetica \\ \hline
training sessions & 351,268 & 433,643      & 719,470    \\
test sessions     & 25,898  & 15,132       & 60,858     \\
\# of items        & 40,728  & 36,968       & 43,097     \\
average lengths   & 6.69    & 5.43         & 5.12       \\ \hline
\end{tabular}
\vspace{0.35cm}
\caption{Statistics of datasets used in experiments.}
\label{Table: Statistics}
\end{table}

\begin{table*}[ht]
\setlength{\tabcolsep}{1.5pt}
\renewcommand{\arraystretch}{1.33}
\caption{The accuracy and long-tail comparison of HID and HID (w/o attr.) which replace attributes of items by semantic clusters.}
\scalebox{0.85}{
\begin{tabular}{cc|cccccc|cccccc|cccccc}
\hline
\multicolumn{2}{c|}{Datasets}                                                       & \multicolumn{6}{c|}{Tmall}                                                                                                                                                                         & \multicolumn{6}{c|}{Diginetica}                                                                                                                                                                             & \multicolumn{6}{c}{Retailrocket}                                                                                                                                                                           \\ \hline
\multicolumn{2}{c|}{Metrics}                                                        & \multicolumn{2}{c|}{Accuracy}                                   & \multicolumn{4}{c|}{Long-tail}                                                                                                   & \multicolumn{2}{c|}{Accuracy}                                   & \multicolumn{4}{c|}{Long-tail}                                                                                                            & \multicolumn{2}{c|}{Accuracy}                                   & \multicolumn{4}{c}{Long-tail}                                                                                                            \\ \hline
\multicolumn{1}{c|}{SBR Models}             & Comparisons                           & HR                        & \multicolumn{1}{c|}{MRR}            & tHR                       & tMRR                      & tCov                               & Tail                                & HR                        & \multicolumn{1}{c|}{MRR}            & tHR                                & tMRR                      & tCov                               & Tail                                & HR                        & \multicolumn{1}{c|}{MRR}            & tHR                       & tMRR                               & tCov                               & Tail                               \\ \hline
\multicolumn{1}{c|}{\multirow{2}{*}{STAMP}} & + HID                                 & \textbf{28.26}            & \multicolumn{1}{c|}{\textbf{15.84}} & \textbf{28.35}            & \textbf{15.93}            & 73.65                              & 78.19                               & \textbf{50.39}            & \multicolumn{1}{c|}{\textbf{17.58}} & \textbf{48.09}                     & \textbf{17.28}            & 93.05                              & 69.24                               & \textbf{52.38}            & \multicolumn{1}{c|}{27.99}          & \textbf{52.09}            & 28.34                              & 56.02                              & 72.59                              \\
\multicolumn{1}{c|}{}                       & \multicolumn{1}{l|}{+ HID (w/o attr.)} & \multicolumn{1}{l}{28.03} & \multicolumn{1}{l|}{15.60}          & \multicolumn{1}{l}{28.08} & \multicolumn{1}{l}{15.76} & \multicolumn{1}{l}{\textbf{73.71}} & \multicolumn{1}{l|}{\textbf{78.29}} & \multicolumn{1}{l}{50.12} & \multicolumn{1}{l|}{17.44}          & \multicolumn{1}{l}{47.79}          & \multicolumn{1}{l}{17.13} & \multicolumn{1}{l}{\textbf{93.17}} & \multicolumn{1}{l|}{\textbf{69.31}} & \multicolumn{1}{l}{52.24} & \multicolumn{1}{l|}{\textbf{28.09}} & \multicolumn{1}{l}{51.97} & \multicolumn{1}{l}{\textbf{28.42}} & \multicolumn{1}{l}{\textbf{56.38}} & \multicolumn{1}{l}{\textbf{72.71}} \\ \hline
\multicolumn{1}{c|}{\multirow{2}{*}{SRGNN}} & + HID                                 & \textbf{28.38}            & \multicolumn{1}{c|}{\textbf{14.66}} & \textbf{28.13}            & \textbf{14.50}            & 66.40                              & 78.12                               & \textbf{52.09}            & \multicolumn{1}{c|}{\textbf{18.26}} & 49.79                              & \textbf{17.25}            & 96.22                              & 70.05                               & \textbf{53.45}            & \multicolumn{1}{c|}{29.47}          & \textbf{52.61}            & 29.51                              & \textbf{55.75}                     & \textbf{73.54}                     \\
\multicolumn{1}{c|}{}                       & \multicolumn{1}{l|}{+ HID (w/o attr.)} & \multicolumn{1}{l}{28.22} & \multicolumn{1}{l|}{14.51}          & \multicolumn{1}{l}{28.00} & \multicolumn{1}{l}{14.36} & \multicolumn{1}{l}{\textbf{66.71}} & \multicolumn{1}{l|}{\textbf{78.31}} & \multicolumn{1}{l}{52.01} & \multicolumn{1}{l|}{18.19}          & \multicolumn{1}{l}{\textbf{49.82}} & \multicolumn{1}{l}{17.18} & \multicolumn{1}{l}{\textbf{96.39}} & \multicolumn{1}{l|}{\textbf{70.18}} & \multicolumn{1}{l}{53.33} & \multicolumn{1}{l|}{\textbf{29.55}} & \multicolumn{1}{l}{52.46} & \multicolumn{1}{l}{\textbf{29.59}} & \multicolumn{1}{l}{55.68}          & \multicolumn{1}{l}{73.48}          \\ \hline
\end{tabular}}
\end{table*}

Note that these are not comparison targets for HID, but rather base models that integrate with HID. Therefore, we have selected the representative high-citated SBR models to demonstrate the generalizability of HID. For comparisons, we select some long-tail approaches which are also plugins:
\begin{itemize}
    \item \textbf{TailNet}~\cite{Liu2020TailNet} is the first classical work in SBR to consider recommendation diversity through the preference mechanism to adjust the importance of tail and head items.
    \item \textbf{CSBR}~\cite{Chen2023CSBR} addresses the long-tail issue of recommendations with two additional training objectives including the distribution prediction and distribution alignment.
    \item \textbf{LOAM}~\cite{Yang2023LOAM} address the long-tail issue of recommendation results through the niche-walk augmentation and the tail session mixup.
    \item \textbf{LAP-SR}~\cite{Peng2024LAPSR} is a post-processing approach that aims  to alleviate the long-tail impact in session-based recommender systems by using  personalized diversity.
\end{itemize}

Since HID focuses on addressing the long-tail issue, \textit{existing intent-based SBR models~\cite{Chen22Intent, Choi2024Intent,Zhang23Efficiently,Wang2024HearInt} which only concentrate on the accuracy of recommendations does not serve as the competitors}. Besides, Since MELT~\cite{Kim2023MELT} , GUME~\cite{Lin2024GUME},  GALORE~\cite{Luo2023Improving}, and LLM-ESR~\cite{Liu24LLMESR} utilize collaborative signals from users, which are not available in session-based recommendation due to the anonymity, we have not included them in the competitors either.

\subsection*{D.4~~~Metrics.}
To evaluate the recommendation accuracy and long-tail performance, we employ three widely used accuracy metrics, which are HR@K, and MRR@K. Following previous works on long-tail issue~\cite{Himan19Managing, Liu2020TailNet, Yang2023LOAM}, we introduce some long-tail metrics which are tHR@K, tMRR@K, tCov@K, and Tail@K. tNDCG@K, HRt@K, and MRRt@K calculate the Normalized Discounted Cumulative Gain, Hit Ratio, and Mean Reciprocal Rank of sessions whose next-item (i.e, ground turth item) belongs to the tail items. For the other two long-tail metrics, we give clear definitions as follows:

\textbf{tCov@K} (Tail Coverage)~\cite{Liu2020TailNet, Yang2023LOAM} measures how many different tail items ever appear in the top-K recommendations, which can be formulated as: ~$tCov@K = \frac{|\cup_{u\in U} L^T_K(u)|}{|V|}$, where $L^T_K(u)$ is the set of long tail items within the top-K recommendations of session $u$.

\textbf{Tail@K}~\cite{Liu2020TailNet, Yang2023LOAM} measures how many long-tail items in the top-K
 for each recommendation list. This metric can be formulated as:~$Tail@K=\frac{1}{|U|}\sum_{u \in U}\frac{|L_K^T(u)|}{K}$, where $U$ is the set of sessions.

\subsection*{D.5~~~Implementation Details}
For general settings, the embedding size is 100, the batch size is 256 for Tmall and Diginetica. The scale parameter $\epsilon$ is set to 0.2, while the temperature coefficient $\sigma$ is set to 0.14. Additionally, the penalty threshold $\eta$ is set to 0.2 and the penalty scale $\lambda_p$ is set to 0.3. For the hybrid intent learning, the number of clusters $n$ of the spectral clustering is set to 300. About the training process, we adopt the Adam optimizer and set the initial learning rate and $L_{2}$ regularization to be 0.001 and $10^{-5}$, respectively, and utilize a StepLR scheduler whose decay rate is 0.6 for each epoch to schedule the learning rate. Considering the training epoch, we set the maximum number of epochs to 20, and stopped training when the model did not show any performance improvement after 3 epochs.

All the parameters are initialized by sampling from a Gaussian distribution. Apart from the above settings, we adopt the best hyperparameters reported in the original papers for all SBR models and comparison methods.

\subsection*{D.6~~~Replacing Attribute With Semantic Clusters}
To enhance the applicability of HID across more scenarios, we design experiments to investigate the performance of HID that does not rely on the real attributes of items. Specifically, we replaced the attributes (which depend on additional information) in the original two-stage process with the results of semantic clustering on item embeddings as the preliminary intent, thereby forming another two-stage clustering method (semantic clustering + spectral clustering) to eliminate reliance on attribute labels.

However, this design renders the pre-storage of hybrid intent infeasible under the original approach, as updates to embeddings alter the mapping between items and preliminary intents—consequently affecting the input to spectral clustering. To address this, we first perform spectral clustering with items as nodes to pre-establish the mapping from items to spectral clusters. During training, we compute the semantic clustering of items in each epoch to update the mapping from items to preliminary intents (i.e., semantic clusters). Subsequently, we retrieve the mapping from items to topological clusters, replace items with preliminary intents, and finally obtain the embedding of hybrid intent through average pooling.

The accuracy and long-tail performance of HID and HID (w/o attr.) are given in Table~2. Here we set the semantic cluster of HID (w/o attr.) to be as the same as the attribute number of HID on each dataset. The results show that HID and HID (w/o attr.) achieve comparable overall performance, confirming that HID outperforms base models in both accuracy and long-tail metrics regardless of additional attribute availability. This validates the effectiveness and generalizability of ICLoss. Notably, HID (w/o attr.) exhibits superior long-tail performance, which may due to the initial meaningless item embeddings. Preliminary intents derived from semantic clustering enable HID (w/o attr.) to explore more item combinations at the initial training stage, thereby enhancing long-tail performance.


\end{document}